\documentclass[12pt]{article}

\usepackage[body={17.5cm, 23cm},right=2cm]{geometry}

\usepackage{color}
\usepackage{graphicx}
\usepackage{epsf}
\usepackage{graphicx,epsfig}
\usepackage{amsmath}
\usepackage[mathscr]{euscript}
\usepackage{pbox}
\usepackage{float}
\usepackage{amsmath}
\usepackage{amssymb}
\usepackage{epsfig}
\usepackage{cite}
\usepackage{color,colordvi}
\newcommand{\be}{\begin{eqnarray}}
\newcommand{\ee}{\end{eqnarray}}
\newcommand{\bi}{\begin{itemize}}
\newcommand{\ei}{\end{itemize}}

\newcounter{hran}

\def\MSbar{\relax\ifmmode\overline{\rm MS}\else{$\overline{\rm MS}${ }}\fi}

\def\bz{\bar{z}}
\renewcommand\Re{\operatorname{Re}}
\renewcommand\Im{\operatorname{Im}}
\begin{document}\thispagestyle{empty}

\vspace{0.5cm}

\def\thefootnote{\arabic{footnote}}
\setcounter{footnote}{0}

\def\s{\sigma}
\def\nn{\nonumber}
\def\p{\partial}
\def\ls{\left[}
\def\rs{\right]}
\def\lc{\left\{}
\def\rc{\right\}}
\def\S{\Sigma}
\def\l{\lambda}
\newcommand{\beq}{\begin{equation}}
\newcommand{\eeq}[1]{\label{#1}\end{equation}}
\newcommand{\bea}{\begin{eqnarray}}
\newcommand{\eea}[1]{\label{#1}\end{eqnarray}}

\renewcommand{\be}{\begin{eqnarray}}
\renewcommand{\ee}{\end{eqnarray}}
\renewcommand{\th}{\theta}
\newcommand{\bth}{\overline{\theta}}

\hspace*{12cm}
CERN-PH-TH/2013-234

\vspace*{2cm}

\begin{center}

{\Large \bf 
Vacuum structure in a chiral ${\cal R}+{\cal R}^n$ modification  of pure supergravity
 %\\ [0.4cm]  and 
 }
\\[1.5cm]
{\normalsize   Sergio Ferrara$^{1,2,*}$,  Alex Kehagias$^{3,4}$ and Massimo Porrati$^{5}$
}
\\[1.1cm]

\vspace{.1cm}
{\small {  $^{1}$ Physics Department, Theory Unit, CERN,
CH 1211, Geneva 23, Switzerland}}\\

\vspace{.1cm}
{\small {  $^{2}$ INFN - Laboratori Nazionali di Frascati,
Via Enrico Fermi 40, I-00044 Frascati, Italy}}\\

\vspace{.3cm}
{\small { $^{3}$ Physics Division, National Technical University of Athens, 15780 Zografou Campus, 
Athens, Greece}}\\

\vspace{.1cm}
{\small {  $^{4}$ Department of Theoretical Physics
24 quai E. Ansermet, CH-1211 Geneva 4, Switzerland}}\\

\vspace{.1cm}
{\small {  $^{5}$
CCPP, Department of Physics, NYU 4 Washington Pl. New York NY 10016, USA}}

%\vspace{.3cm}

%\vspace{.2cm}

\end{center}

\vspace{1.2cm}

%\hrule \vspace{0.3cm}
\begin{center}
{\small  \noindent \textbf{Abstract}} \\[0.5cm]
\end{center}
\noindent 
{\small
We discuss an ${\cal R}+{\cal R}^n$ class of modified ${\cal N}=1$, $D=4$ supergravity models  
where the deformation is a monomial ${\cal R}^n\big|_F$ in the 
 chiral scalar curvature multiplet ${\cal R}$ of the ``old minimal'' auxiliary field formulation. The scalaron and goldstino multiplets are dual to each other in this theory.
 Since one of them is not dynamical, this theory, as recently shown, cannot be used 
 as the supersymmetric completion of $R+R^n$ gravity.
This is confirmed by investigating the scalar potential and its critical points in the dual standard supergravity 
formulation with 
a single chiral  multiplet with specific K\"ahler potential and  superpotential.
We study the vacuum structure of this dual theory and we find that there is always a supersymmetric  
 Minkowski critical point which however is pathological for $n\geq 3$ as it corresponds to a corner ($n=3$) and a cusp
 ($n>3$) point of the potential. For $n>3$ an  anti-de Sitter regular supersymmetric vacuum 
emerges. As a result, this class of models are not appropriate to describe  inflation. 
We also find the mass spectrum and we provide a general formula for the masses of the scalars of a
chiral multiplet around the anti-de Sitter critical point and their relation to $osp(1,4)$ unitary representations.   
}
%\vspace{0.5cm}  \hrule
\vskip 2cm

\def\thefootnote{\arabic{footnote}}
\setcounter{footnote}{0}

\vskip.2in
\line(1,0){250}\\
{\footnotesize {$^*$On leave of absence from Department of Physics and Astronomy, University of California Los Angeles, CA 90095-1547
USA}}

%\xfilll[-12pt]{12pt}

%\maketitle

%\date{\today}

\baselineskip= 19pt
\newpage 

\section{Introduction}

Motivated by the latest Planck mission data \cite{planck1,planck2}, there has 
been recently a renewed interest 
in $R+R^2$ bosonic theories, 
which realize the Strarobinsky model of inflation. The supersymmetric extension of such theories depends on the off-shell 
degrees of freedom of supergravity.  As there are two such minimal extensions, old and new minimal supergravity,
there are two inequivalent ways to supersymmetrize the bosonic $R+R^2$ theory. 
This has been done originally  in \cite{cecotti} in the old minimal supergravity
framework and in \cite{CFPS} in new minimal formulation. 
These theories have the common feature of adding to pure supergravity four bosonic and four fermionic degrees of freedom,
two chiral multiplets in old minimal \cite{cecotti} and a massive vector multiplet in new minimal \cite{CFPS},
in accordance with the linearized analysis given in \cite{FGN}.
Recently, these theories have been considered 
\cite{KL,KL1,EON,EON1,FKR,FKLP12,Fre} in the light of the new constraints set by the Planck mission on inflation \cite{planck2}. 
 
In the same spirit, there was an effort to supersymmetrize bosonic $f(R)$ gravity theories, called 
 ``$F(R)$ supergravity'' \cite{KS0,KS1}. On the gravity side, one may eliminate either the vector auxiliary $A_\mu$ 
 or the complex scalar
 auxiliary $X$ of the off-shell gravity multiplet depending on an integration by parts.
 Integrating out $A_\mu$ one gets a gravity theory with propagating $X$ \cite{FKvP}. 
 %This theory is dual to standard supergravity with 
 %a chiral multiplet $\Lambda|_{\theta=0}=f'(R)$. 
 Integrating $X$, one gets a non-linear theory for $R$ and $A_\mu$
 where both the scalaron and  $D^\mu A_\mu$ are propagating.
 This theory when $A_\mu$ is neglected reduces to theory considered in \cite{KS0,KS1}.
 Both supergravities are dual to standard supergravity coupled to 
 a  chiral multiplet. The latter is dual to two  propagating bosonic degrees of freedom on the supergravity side, 
 depending on which one of
 the auxiliary fields has been 
 integrated out. Here we will investigate the dual theory  for 
models with higher powers of the chiral superfield curvature multiplet.  
We will see that for this class of models, the dual theory is standard supergravity 
coupled to a single chiral multiplet
with a no-scale K\"ahler potential and a given  superpotential. The induced scalar potential fails to have a de Sitter 
asymptotic regime as it has already been observed in \cite{EON1}.
In addition, we will find the vacuum structure of the resulting
supergravity and we will provide a general formula for the masses of the scalars on supersymmetric AdS vacuua. 

Theories without two gravitationally generated chiral multiplets fail to reproduce the linearized analysis of 
\cite{FGN} and, as shown in \cite{FKvP},
$R+R^2$ or  $R^n$  power modifications of Einstein supergravity.
Models which are the supersymmetric completion of $R+R^2$ and reproduce the 
Starobinsky model
are not unique since they allow a 
K\"ahler potential and superpotential for the goldstino multiplet \cite{cecotti}.
Actually, as shown  in \cite{KL1} and \cite{FKvP}, non minimal K\"ahler potential modifications for the 
goldstino are required for
the inflaton flow to be a stable direction in field space. The same class of models have been recently 
revisited and further investigated by the proponents of $F(R)$ supergravity \cite{ketov2}.

In the next section 2, we review modifications of gravity by higher curvature terms in the old minimal formulation. 
In section 3,
we discuss particular ${\cal R}+{\cal R}^n$ modifications and the structure of the vacuum of the dual standard 
supergravity. 
In section 4, we calculate the mass spectrum of a supergravity theory coupled to a single chiral multiplet around a
supersymmetric AdS vacuum and we identify it with unitary representations of $osp(1,4)$. Finally,  we conclude in section 5.

\section{Modified Supergravity By Higher Curvature Terms}

The $R+R^2$ theory, as is it revealed by 
a linearized analysis \cite{FGN} contains the degrees of freedom of 
two chiral multiplets. 
Three of them come from the Einstein supergravity auxiliaries $X,\partial^\mu A_\mu$ where \cite{FvN,SW,FvP}

\be
X=\frac{1}{3}u=\frac{1}{3}(S-iP)\, , 
\ee
and the fourth comes from the 
scalar curvature R. The minimal $R+R^2$ theory is given by 

\be
{\cal L}_{R+R^2}=\gamma [S_0\bar{S}_0]_D+\alpha [{\cal R}\bar {\cal R}]_D ,
\ee
where the two D-terms above are the supersymmetric extension of the $R+R^2$ bosonic theory.
Here $S_0$ is the compensator chiral superfield, with scaling weight and chiral 
weight equal to 1, the
curvature chiral superfield ${\cal R}$  has scaling and chiral weight equal to 1 as well, 
and $[O]_{D,F}$ are the
standard D- and F-term density formulae of conformal supergravity, where $O$ is a real 
superfield with scaling
weight 2 and vanishing chiral weight. 
The bosonic components of the  curvature 
chiral scalar multiplet  ${\cal R}$  are

\be
{\cal R}=\bar X +\cdots+\theta^2 {\mathscr{F}}_R,
\ee
where 
\be
{\mathscr{F}}_R=-\frac{1}{2} R-3 A_\mu^2+3 i D^\mu A_\mu .
\ee
Let us also note that ${\cal R}/S_0$ is of zero chiral and Weyl weight and its bosonic content is 
\be
{\cal R}/S_0=\bar X+\cdots +\theta^2 ({\mathscr{F}}_R-18 X\bar X).
\ee

There is an alternative modified supersymmetric action, considered in the literature \cite{KS0,KS1} as an alternative 
``f(R)'' action, given by the F-term

\be
{\cal L}_{f(R)}=[F\big({\cal R}/S_0\big)S_0^3]_F ,  \label{ff}
\ee
whose linear and constant terms are representing the Einstein term and a cosmological constant. 
The other higher order terms make only one of the two chiral multiplet degrees of freedom propagating and therefore cannot 
describe $R+R^2$ gravity.
As shown in \cite{FKvP}, the bosonic part of this action (including all auxiliary fields $X,A_\mu$) is

\be
{\cal L}_{bos}=-\frac{1}{2} \sqrt{-g}\Big\{
27 X F(\bar X)-18 F'(\bar X)X \bar X+F'(\bar X) {\mathscr{F}}_R\Big\}+h.c. \, .
\label{6}
\ee
By noticing that 
$
\Im {\mathscr{F}}_R=3\partial^\mu A_\mu$
and integrating by parts this term we can solve for $A_\mu$,
%$\delta {\cal L}/\delta A_\mu=0$
%\cite{FKvP}. 
Then in 
eq.~(\ref{6}) we find that the scalar $X$ is propagating \cite{FKvP}

\begin{align}
{\cal L}_{bos}\Big|_{\delta {\cal L}/\delta A_\mu=0}&=
\frac{1}{2}\sqrt{-g} R-\frac{3}{4}\frac{1}{[\Re F'(\bar{X})]^2}
\Big\{\big(\partial_\mu \Re F'(\bar{X})\big)^2+\big(\partial_\mu \Im F'(\bar{X})\big)^2\Big\}\nonumber \\
&+\sqrt{-g}\Big(-27 \Re(X F(\bar{X}))+18 \Re \big(F'(\bar{X})\big)   X\bar X\Big) . \label{8}
\end{align} 

This Lagrangian has a gravity dual with a seemingly different action involving  non-linear $R$ terms and the 
propagating $\partial^\mu A_\mu$ 
auxiliary scalar. The dual action is obtained by integrating $X$ in 
eq.~(\ref{6}) and leads to
\begin{align}
{\cal L}_{bos}\Big|_{\delta {\cal L}/\delta X=0}&={\cal L}_{bos}^D({\mathscr{F}}_R,\bar{{\mathscr{F}}}_R). \label{9}
\end{align}
If we set $A_\mu=0$ so that ${\mathscr{F}}_R=-R/2$, we get the non-linear R-theory constructed in \cite{KS0,KS1}.
%Since this complicated theory when $A_\mu$ is not neglected, is
The dual gravity theories described by eq.(\ref{8},\ref{9}) are both dual to a standard supergravity of a self interacting 
chiral multiplet with a superpotential term.  We are going to investigate 
the vacuum of the latter for a particular class of models where the superpotential can easily be computed.

From eq.(\ref{8}) it is obvious that the physical chiral multiplet is $\Lambda=F'(\bar{X})$ so 
a Legendre transform can be 
performed to express the theory in $\Lambda$ (rather that $\bar{X}$) 
variables, and to find its  superpotential.   To explicitly show this, 
 one may consider an action  in superconformal calculus and in the old minimal supergravity context, of the form

\be
{\cal L}=-[S_0\bar{S}_0]_D+[S_0^3 f({\cal R}/S_0)]_F . \label{a1}
\ee
This theory can be obtained from
\be
{\cal L}_D=-[S_0\bar{S}_0]_D+[\Lambda \left(A-\frac{{\cal R}}{S_0}\right)S_0^3]_F-
[S_0^3 f(A)]_F .  \label{a2}
\ee
Indeed, integrating out the Lagrange multiplier superfield $\Lambda$ in (\ref{a2}), we get back the original theory 
(\ref{a1}).
However, by using the identity \cite{cecotti,FKvP}
\be
[\Lambda R S_0^2]_F=[(\Lambda+\bar \Lambda)S_0\bar{S}_0]_D ,
\ee
we may write (\ref{a2}) as
\be
{\cal L}_D=-[(1+\Lambda+ \bar \Lambda)S_0\bar{S}_0]_D+[ \left(\Lambda A-f(A)\right)S_0^3]_F .\label{a3}
\ee
By integrating out the chiral Lagrange multiplier $A$, we obtain the dual action
\be
{\cal L}_D\Big|_{\delta{\cal L}_D/\delta A}=-[(1+\Lambda + \bar \Lambda)S_0\bar{S}_0]_D+ [W(\Lambda)S_0^3]_F ,
\ee
where 
\be
W(\Lambda)=\big(Af'(A)-f(A)\Big)\Big|_{f'(A)=\Lambda} .
\ee

\section{${\cal R}^n$ Modification of Supergravity}

As we have seen above, the theory (\ref{ff}) can be described in a dual formulation by standard supergravity coupled to 
a single chiral multiplet. 
Although the discussion could be kept general, we will consider here the case
\be
f(A)=\varepsilon_n\, A^n ,
\ee
which corresponds to the choice $(F=-1-f)$
\be
F({\cal R}/S_0)=-{\cal R}/S_0+\epsilon_n ({\cal R}/S_0)^n .
\ee
In this case we get the superpotential
\be
W(\Lambda)= \lambda_n\, \Lambda^{\frac{n}{n-1}}\, , ~~~\lambda_n=\varepsilon_n^{\frac{1}{1-n}}(n-1)n^{\frac{n}{1-n}}
\ee
and after defining $\Lambda$ as
\be
\Lambda=T-\frac{1}{2},
\ee
the theory is described by 
\be
{\cal L}=-[(T+\bar T)S_0\bar{S}_0]_D+\lambda_n [\Big(T-\frac{1}{2}\Big)^{\frac{n}{n-1}}S_0^3]_F\, .
\ee
In other words, the theory has been turned into standard supergravity with a no-scale K\"ahler potential 
\cite{ns1}
\be
K=-3 \log(T+\bar T) , ~~~\Re T>0   \label{w}
\ee
and superpotential 
\be
W(T)=\lambda_n\left(T-\frac{1}{2}\right)^\frac{n}{n-1}=\frac{\lambda_n}{2^\frac{n}{n-1}}\left(C-1+iB\right)^\frac{n}{n-1}\, ,
\ee
where we have parametrized $T$ as $T=(C+iB)/2$ so $C>0$. 
It is straightforward now to calculate the potential using the standard formula

\be
V=\frac{1}{(T+\bar T)^2}\left\{ \frac{1}{3} (T+\bar T) |W_T|^2-W \bar{W}_T-\bar{W} W_T\right\}\, .
\ee
Explicitly, we have 

\be
V=\frac{\lambda_n^2}{(T+\bar T)^2}\frac{n}{n\!-\!1}\Big|T-\frac{1}{2}\Big|^{\frac{2}{n-1}}
\left\{\frac{n}{n\!-\!1}\frac{T+\bar T}{3}-(T+\bar T)+1\right\} ,
\ee
%which for
% %\be
%T=\frac{1}{2}\Big(C+iB\Big)
%\ee
and in terms of $C,B$, the potential $V$ is 
%turns out to be

\be
V=\frac{\lambda_n^2}{4^{\frac{1}{n\!-\!1}}}\frac{n}{n\!-\!1}\frac{1}{C^2}\left\{\phantom{\frac{X}{Y}}\!\!\!\!\!\!\!
(C-1)^2+B^2\right\}^{\frac{1}{n-1}}
\left\{C \, 
\frac{3\!-\!2n}{3(n\!-\!1)}+1\right\} \, . \label{v}
\ee
The form of the potential for $n=2,3$ and $n>3$ has been plotted in Figure 1. Note that the canonically normalized scalar
is $\phi$ defined by $C=2\Re T=e^{\sqrt{\frac{2}{3}}\phi}$. 
\begin{figure}
 \includegraphics[width=0.5\textwidth]{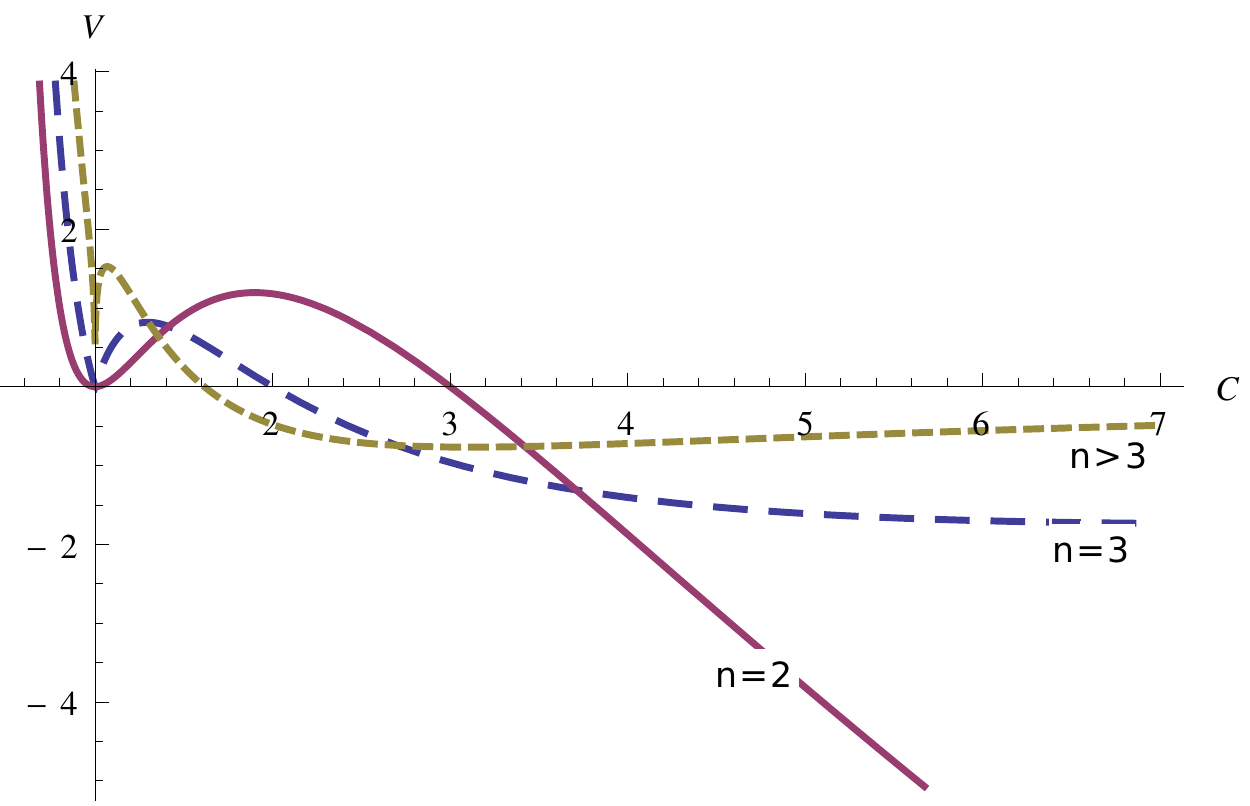}
 \includegraphics[width=0.5\textwidth]{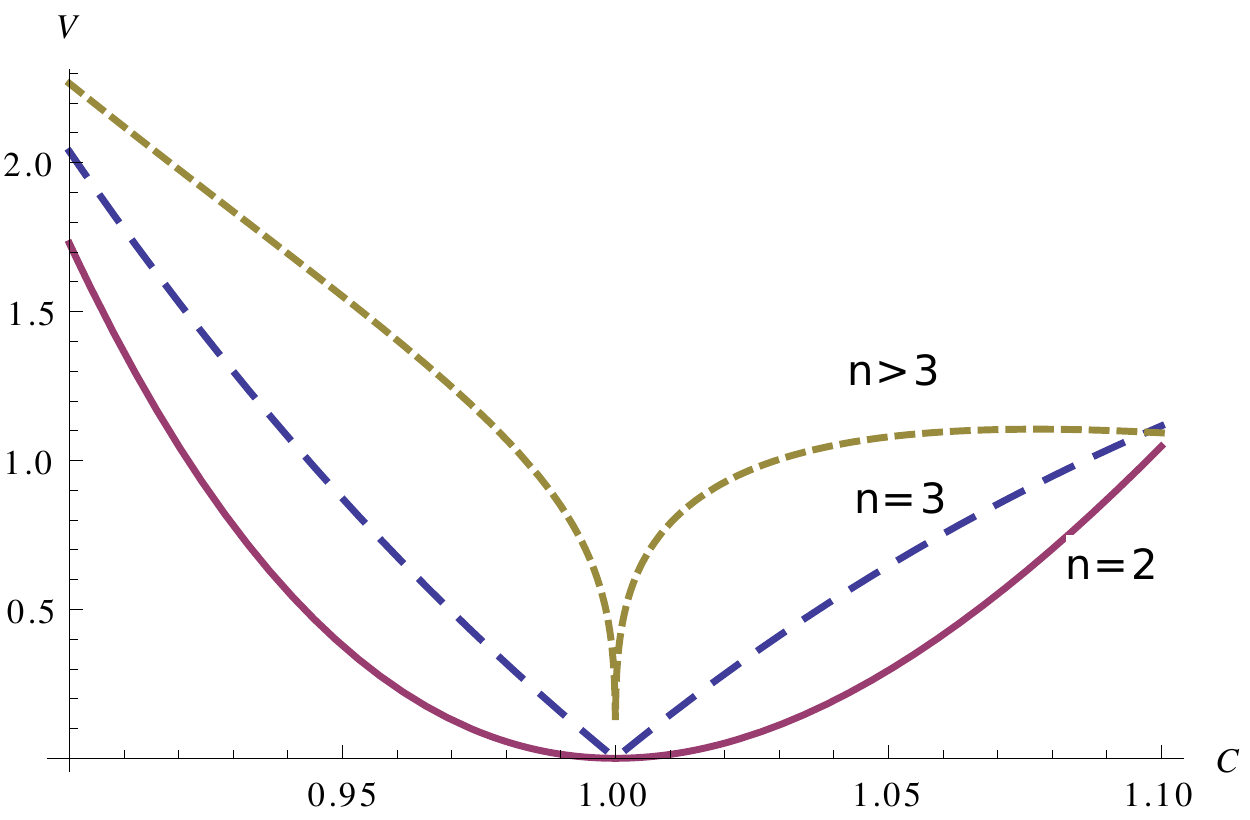}
 \caption{The scalar potential $V(C)$ at $B=0$ and $n=2,3 $ and $n>3$ is given 
in the left figure. The right figure magnifies the region near $C=1$, 
where the ``corner'' and the cusp are easily recognized for $n=3$ and $n>3$,
 respectively. }
\end{figure}

%It is straigtforward to find the supersymmetric critical points of the potential. 
In order to find the supersymmetric vacua  of the theory, one should look for the solutions of 
\be
D_TW=\partial_TW+K_T W=0, \label{s}
\ee
which  are $B=0$ and 

\be
\frac{1}{2^{\frac{1}{n-1}}}\Big(C-1\Big)^{\frac{1}{n-1}}\left\{\frac{n}{n\!-\!1}-\frac{3}{2}
\frac{1}{C}\big(C-1\big)\right\}=0 \, .\label{s1}
\ee
Eq.(\ref{s1}) 
has two solutions:
\be
W=\partial_TW=0\, , ~~~(V=0), ~~~~C=1, ~~B=0 \, , ~~~ \mbox{(Minkowski)}, \label{sm}
\ee
and 
%which corresponds to a supersymmetric Minkowski vacuum and 
%The second solution is provided by eq.(\ref{s}) with 
\be
W\neq 0, ~~~(V=-3 e^{G}) ~~~ G=K+\log|W|^2\, ,  ~~~\mbox{(AdS)}. \label{s2}
\ee
%which correspond to a supersymmetric anti-de Sitter vacuum. 
A  solution to eq.~(\ref{s2}) exists for $n>3$ and it is explicitly given by 

\be
C=\frac{3(n-1)}{n-3}\, , ~~B=0,~~~~~n>3,  ~~~~\mbox{AdS vacuum}.  \label{anti}
\ee
Note that the zeros  of the potential $V=0$ are 
at 
\be
C=1, ~~B=0 ~~~\mbox{and}~~~C=\frac{3(n-1)}{2n-3} , ~~B=0.
\ee

%There are two solutions to this equation
%\be
%&&a)~~C=1 \, ,~~~~\mbox{Minckowski vacuum}\\
%&&b)~~C=\frac{3(n-1)}{n-3}\, , ~~n>3,  ~~~~\mbox{AdS vacuum}
%\ee

%the extrema of the potential (\ref{v}). 
In order to explicitly study  the vacuum structure of the theory,  we should distinguish  three  
cases according to the 
asymptotic behaviour of the potential for large values of the fields $C$. As $ C\to \infty$ we may have: 
I) $V\to -\infty$, II) $V\to -\frac{3}{8}$ and III) $V\to 0^{-}$. These cases correspond to: 
I) $n=2$, II) $n=3$ and III) $n>3$, respectively. 
In the  case I), there exists a local minimum at $C=1$ where the potential vanishes and corresponds,
as we will see, to a supersymmetric 
Minkowski vacuum. 
There is also a maximum at $C=2$ which is not supersymmetric. 
In the $n=3$ case, there exists  a supersymmetric Minkowski vacuum at $C=1$, which is now a 
``corner'' i.e. a point where the first derivative has a 
finite discontinuity.  
There  exists also a non-supersymmetric maximum which is at $C=4/3$. Finally, for $n>3$ there exists the 
supersymmetric Minkowski 
vacuum at $C=1$, which is now a cusp, the non-supersymmetric maximum at 

\be
C_1=\frac{2(n-1)}{2n-3}\, , ~~~B_1=0 ,
\ee
 and a supersymmetric one  at 
\be
C_2=\frac{3(n-1)}{n-3}\, , ~~~B_2=0 .
\ee
The above vacuum structure has been tabulated in  Table 1.

\begin{table}[H]
\renewcommand{\arraystretch}{1.3}
\begin{center}
    \begin{tabular}{|l||l|l|}
    \hline \hline
    $n$   & SUSY mimima       $V\leq 0$                   & non-SUSY maxima $V>0 $                \\ \hline \hline
    $n=2$ & $C=1$,       Minkowski                          & $C=2$           \\ \hline
    $n=3$ & $C=1$,  Minkowski (corner)                         & $C=\frac{4}{3}$          \\ \hline
    $n>3$ & \pbox{20cm}{$C=1$, Minkowski (cusp), \\  $C=\frac{3(n-1)}{(n-3)}$, $AdS_4$} & $C=\frac{2(n-1)}{2n-3}$\\ 
    \hline \hline
    \end{tabular}
\caption{Supersymmetric and non-supersymmetric critical points of the potential V.}  
    \end{center}
\end{table}

The masses of the $C,B$ fields are given in the two cases as (with $\lambda_n=1$)
\be
&&m_{C_1}^2=2 K_{T\bar{T}}^{-1} V_{CC}\Big|_{C_1,B_1}=
-\frac{2^{\frac{2}{1-n}} n (4 n-3)(2 n-3)^{\frac{2(n-2)}{n-1}} }{9 (n-1)^3}<0,\\
%-\frac{ n(3-2n)^{\frac{4n-5}{n-1}}(4n-3)\\
&&m_{B_1}^2=2 K_{T\bar{T}}^{-1} V_{BB}\Big|_{C_1,B_1}=
\frac{2^{\frac{2}{1-n}} n (2 n-3)^{\frac{2(n-2)}{n-1}}}{9(n-1)^2}>0,
%\frac{2^{\frac{n+1}{1-n}}}{3(n-1)^4}n(3-2n)^{\frac{4n-5}{n-1}}
\ee
and 
\be
&&m_{C_2}^2=2 K_{T\bar{T}}^{-1} V_{CC}\Big|_{C_2,B_2}=
\frac{n^{\frac{2}{n-1}}}{27(n-1)^3} (4n-3) (n-3)^{\frac{2(n-2)}{n-1}}>0,\label{m1} \\
&&m_{B_2}^2=2 K_{T\bar{T}}^{-1} V_{BB}\Big|_{C_2,B_2}=-\frac{n^{\frac{2}{n-1}}}{3(n-1)^2}(n-3)^{\frac{n-2}{n-1}}<0 .
\label{m2}
\ee
Note that we have multiplied the second derivatives of the potential by $2 K_{T\bar{T}}^{-1}$ in order 
to canonically normalize the kinetic terms of $C,B$.

%The critical point (a) corresponds to a Minkowski vacuum since at this point $W=\partial_T W=0$.
We should mention here that  the $C=1$ Minkowski point (\ref{sm}) 
 is quite particular. Namely, although it is a normal local minimum for $n=2$, it is a point with discontinuous 
 first derivative (corner) for $n=3$ and singular first derivative for $n>3$ (cusp). This can explicitly be 
 seen in Figure 1, where
 the potential is depicted around $C=1$ for the three cases. As a result, the scalar equations are not satisfied 
 at this point for $n\geq 3$ although  it is a supersymmetric critical point. The values of the potential $V$ and its first 
 derivative $V_C$ at the supersymmetric points have been tabulated in the following Table 2.
\begin{table}[H]
\renewcommand{\arraystretch}{1.6}
\begin{center}
\begin{tabular}{|l|c|c|c|c|}
\hline \hline
      & \multicolumn{3}{c|}{Minkowski}  & AdS                                     \\ \hline \hline
      & $n=2$ & $n=3$           & $n>3$      & $n>3$                                     \\ \hline \hline
V     & 0   & 0             & 0        & -$\frac{n^{\frac{2n}{n-1}}}{9(n-1)^3} (n-3)^{\frac{n-3}{n-1}}$ \\ \hline 
$V_C$ & 0   & $\pm\frac{3}{8}$ & $\infty$ & 0    \\ \hline                               
\end{tabular}
\caption{The value of the potential $V$ and its first derivative $V_C$ at the supersymmetric critical points}
\end{center}
\end{table}

 On the other hand, (\ref{anti}) corresponds to an AdS vacuum since at this point $W\neq0$ and $V<0$. 
 Therefore, we expect that  
 (\ref{m1},\ref{m2}) to be the masses of  unitary representations of the $AdS_4$ simple
 superalgebra $osp(1,4)$. To see this, we will discuss in the next section  the more general case of a single 
 chiral multiplet in  supergravity. 
 
 \section{Supersymmetric AdS vacua and masses of unitary multiplets of $osp(1,4)$}
 Let us consider the general form of the ${\cal N}=1$ scalar 
 potential of a single multiplet $z$ 
\be
V=
e^G\Big(G_zG_{\bar{z}} 
G^{-1}_{z\bz}-3\Big) .
\ee
Since
\be
V_z=e^GG_z^2G_{\bz}G^{-1}_{z\bz}+e^GG_{zz}G_{\bz}G^{-1}_{z\bz}+e^G G_z+e^GG_zG_{\bz} (G^{-1}_{zz})_{\bz}-3 e^GG_z ,
\ee
it is easy to see that critical points of the potential are points where 
\be
G_z=G_{\bz}=0 . \label{crit}
\ee
These correspond to supersymmetric $AdS_4$ vacua with cosmological constant 

\be
\Lambda=V\Big|_{G_z=G_{\bz}=0}=-3e^G .
\ee
The $AdS_4$ scalar curvature is $R=-12 L_{AdS}^{-2}$ where the $AdS_4$ radius $L_{AdS}$ is 
\be
L_{AdS}^2=-\frac{3}{\Lambda}= e^{-G}
\ee
and the Breitenlohner-Freedman bound in $d+1$-spacetime dimensions
\be
m^2L_{AdS}^2\geq -\frac{d^2}{4}
\ee
is written in our case as
\be
m^2\geq -\frac{9}{4} ,
\ee
in units of $e^G$.
At the critical points (\ref{crit}), we find that 
\be
V_{zz}=- e^GG_{zz}\, , ~~~~V_{z\bz}=-2e^G G_{z\bz}+e^GG_{zz}G_{\bz\bz}G^{-1}_{z\bz}
\ee
and therefore, after multiplying with $G^{-1}_{z\bz}$ we get 
\be
&&V_{zz}G^{-1}_{z\bz}=-e^G A,\\
 && V_{z\bz}G^{-1}_{z\bz}=e^G(-2+|A|^2) ,
 \ee
 where
 \be
 A=G_{zz} G^{-1}_{z\bz} .
 \ee
In terms of the real and imaginary parts of $z$ ($C=\sqrt{2}\Re z, ~B=\sqrt{2}\Im z$)
we may write
\be
&&e^{-G}\frac{1}{2}\Big(V_{CC}+V_{BB}\Big)=-2 +|A|^2\, , \\
&&e^{-G}\frac{1}{2}\Big(V_{CC}-V_{BB}\Big)=- \Re A\, , \\
&&e^{-G}V_{CB}=-Im A
\ee
and thus, the mass matrix turns out to be, in $e^G$ units
\be
{\cal M}^2=\left(
\begin{array}{cc}
 V_{CC}&V_{CB}\\
 V_{CB}&V_{BB}
\end{array}\right)=\left(
\begin{array}{cc}
 -2+|A|^2-ReA&-ImA\\
 -Im A&-2+|A|^2+ReA
\end{array}
\right) .
\ee
By diagonalizing the mass matrix, we find that the mass  eigenvalues are  ($m_C^2>m_B^2$)
\be
&&m_B^2=(|A|+1)(|A|-2)\, , \label{mm1} \\
&&m_C^2=(|A|-1)(|A|+2), \label{mm2}
\ee
($m_B^2>m_C^2$, eq.(\ref{mm1}) and eq.(\ref{mm2}) are interchanged)
so that 

\be
\begin{array}{ccl}
|A|\geq 2 , && m_B^2\geq 0\, , ~~m_C^2>0 ,\\
1\leq |A|<2,&& m_B^2<0\, , ~~m_C^2\geq 0 ,\\
|A|<1, && m_C^2,m_B^2<0 .
\end{array}
\ee
%In models where $m_B^2>m_C^2$, eq.(\ref{mm1}) and eq.(\ref{mm2}) are interchanged.
Particular values are 
\be
\begin{array}{llll}
|A|=2,&m_B^2=0,& m_C^2=4 ,&\\
|A|=1,& m_B^2=-2,&m_C^2=0,&G_{zz}=G_{\bz\bz}=G_{z\bz}, \\
|A|=0,&m_B^2=-2,&m_C^2=-2,& G_{zz}=0, G=z\bz .
\end{array}
\ee
By introducing 
\be
E_0=|A|+1 ,
\ee
the mass spectrum may be expressed as
\be
&&m_B^2=E_0(E_0-3)\, , \nonumber \\
&&m_C^2=(E_0-2)(E_0+1) . \label{mm}
\ee
We recognize in (\ref{mm}) the masses of the scalars in the Wess-Zumino unitary representation of $osp(1,4)$ 
\be
D(E_0,J)\oplus D(E_0+\frac{1}{2},\frac{1}{2})\oplus D(E_0+1,0) ,
\ee
where $D(E_0,J)$ are unitary representations of $so(2,3)$ with energy $E_0$ and spin $J$.  
 
 It can easily be checked that eqs.~(\ref{m1},\ref{m2}) are given by (\ref{mm1},\ref{mm2}), respectively. In particular, 
the $AdS_4$ vacuum at $C=\frac{3(n-1)}{n-3}$ for $n>3$ has $|A|=(2n-3)/n$. Therefore $1<|A|<2$ in this case and thus
$m_B^2<0,~ m_C^2>0$ in accordance with (\ref{m1},\ref{m2}). Note also that although $m_B^2<0$ we have 

\be
 m_B^2=-\frac{9(n-1)}{n^2}>-\frac{9}{4}\, , ~~~~ n>3
\ee
and thus, the Breitenlohner-Freedman bound is satisfied.

\section{Conclusions}

We have discuss here a particular  “F(R) supergravity” \cite{KS0,KS1}, 
namely the 
$F({\cal R})={\cal R}+{\cal R}^n$ class of supergravity models. We found that 
this theory is not the supersymmetric completion of $R+R^n$, since it does not 
contain two chiral multiplets. In the old minimal formulation, 
such theory contains extra degrees of freedom for $n>2$ \cite{cecotti}.
% In particular,
%or $n=2$, its bosonic 
%sector does not describe Starobinsky $R+R^2$ gravity and, in general, it does not reproduce
%bosonic $R+R^n$ gravity for any $n$.
By introducing 
appropriate Lagrange multiplier chiral superfields, we found the dual theory, 
which describes  
 a single chiral superfield coupled to supergravity with no-scale K\"ahler potential
 and a superpotential term. We discussed the vacuum structure of 
this theory and we found that it has always a supersymmetric Minkowski local minimum for any $n>1$ and 
a  anti-de Sitter  vacuum for $n>3$.  However, only for $n=2$ this local minimum corresponds, strictly speaking, to a vacuum.
The reason is that for $n=3$ this local minimum is a ``corner'' of the potential whereas, for $n>3$, it is a cusp point.
As a result, the second derivative of the potential, which enters the classical equations of motion for the scalar, 
has either a delta-function peak ($n=3$), or, it is not defined at all ($n>3$). This makes the interpretation 
of this point as a vacuum questionable. 
On the other hand, we found that the theory possess a global supersymmetric
Anti de Sitter minimum for $n>3$~\footnote{It is a minimum because all scalar
masses obey the Breitenlohner-Friedmann bound.}. 
We calculated the masses of the 
scalar fluctuations
around the anti-de Sitter vacuum and we found that they agree with the masses of the scalars of the Wess-Zumino
unitary representation of the simple superalgebra $osp(1,4)$.

\vskip.2in

\noindent
{\bf {Acknowledgment}}

\vskip.1in
\noindent
We would  like to thank R. Kallosh, A. Linde and A. van Proeyen for enlighting discussions on the subject of this investigation. 
This research was implemented
under the ARISTEIA Action of the Operational Programme Education and Lifelong Learning
and is co-funded by the European Social Fund (ESF) and National Resources. It is partially
supported by European Union’s Seventh Framework Programme (FP7/2007-2013) under REA
grant agreement n. 329083. S.F. is supported by ERC Advanced Investigator Grant n. 226455
Supersymmetry, Quantum Gravity and Gauge Fields (Superfields). M.P. is supported in part by
NSF grants PHY-0758032 and PHY-1316452.

%%%%%%%%%%%%%%%%%%%%%%%%%%%%%%%%%%%%%%%%%%%%%%%%%%%%%%%%%%%%%%%%%%%%%%%%

\newpage
\begin{thebibliography}{99}

\bibitem{planck1}
  P.~A.~R.~Ade {\it et al.}  [Planck Collaboration],
  {\it {  {\small    Planck 2013 results. XVI. Cosmological parameters,}}}
  arXiv:1303.5076 [astro-ph.CO].
  %%CITATION = ARXIV:1303.5076;%%
  %681 citations counted in INSPIRE as of 29 Sep 2013

\bibitem{planck2}
  P.~A.~R.~Ade {\it et al.}  [Planck Collaboration],
  {\it { {\small   Planck 2013 results. XXII. Constraints on inflation,}}}
  arXiv:1303.5082 [astro-ph.CO].
  %%CITATION = ARXIV:1303.5082;%%
  %209 citations counted in INSPIRE as of 26 Sep 2013

  
  
  \bibitem{cecotti} 
  S.~Cecotti,
  {\it { {\small   Higher Derivative Supergravity Is Equivalent To Standard Supergravity Coupled To Matter. 1.,}}}
  Phys.\ Lett.\ B {\bf 190}, 86 (1987).
  %%CITATION = PHLTA,B190,86;%%
  %30 citations counted in INSPIRE as of 26 Sep 2013
  
  \bibitem{CFPS} 
  S.~Cecotti, S.~Ferrara, M.~Porrati and S.~Sabharwal,
  {\it { {\small   New Minimal Higher Derivative Supergravity Coupled To Matter,}}}
  Nucl.\ Phys.\ B {\bf 306}, 160 (1988).
  %%CITATION = NUPHA,B306,160;%%
  %17 citations counted in INSPIRE as of 26 Sep 2013
  
  
  \bibitem{KL} 
   R.~Kallosh and A.~Linde,
  {\it { {\small   Superconformal generalizations of the Starobinsky model,}}}
  JCAP {\bf 1306}, 028 (2013)
  [arXiv:1306.3214 [hep-th]].~
  %%CITATION = ARXIV:1306.3214;%%
  %19 citations counted in INSPIRE as of 27 Sep 2013

  \bibitem{KL1}
  R.~Kallosh and A.~Linde,
  {\it { {\small   Universality Class in Conformal Inflation,}}}
  JCAP {\bf 1307}, 002 (2013)
  [arXiv:1306.5220 [hep-th]]. ~
  R.~Kallosh and A.~Linde,
  {\it { {\small   Non-minimal Inflationary Attractors,}}}
  arXiv:1307.7938 [hep-th].~
  %%CITATION = ARXIV:1307.7938;%%
  %4 citations counted in INSPIRE as of 27 Sep 2013 
  R.~Kallosh and A.~Linde,
  {\it { {\small   Multi-field Conformal Cosmological Attractors,}}}
  arXiv:1309.2015 [hep-th].
  %%CITATION = ARXIV:1309.2015;%%
   
  %%CITATION = ARXIV:1306.5220;%%
  %16 citations counted in INSPIRE as of 27 Sep 2013
 
 
 
  \bibitem{EON} 
  J.~Ellis, D.~V.~Nanopoulos and K.~A.~Olive,
  {\it { {\small   No-Scale Supergravity Realization of the Starobinsky Model of Inflation,}}}
  arXiv:1305.1247 [hep-th]. ~  D.~Croon, J.~Ellis and N.~E.~Mavromatos,
  {\it { {\small   Wess-Zumino Inflation in Light of Planck,}}}
  Physics Letters B {\bf 724}, , 165 (2013)
  [arXiv:1303.6253 [astro-ph.CO]].
  %%CITATION = ARXIV:1303.6253;%%
  %8 citations counted in INSPIRE as of 27 Sep 2013
  %%CITATION = ARXIV:1307.3537;%%
  %9 citations counted in INSPIRE as of 27 Sep 2013
  %%CITATION = ARXIV:1305.1247;%%
  %20 citations counted in INSPIRE as of 26 Sep 2013
  
  \bibitem{EON1}
  J.~Ellis, D.~V.~Nanopoulos and K.~A.~Olive,
  {\it { {\small   Starobinsky-like Inflationary Models as Avatars of No-Scale Supergravity,}}}
  arXiv:1307.3537 [hep-th].~ 
 
  \bibitem{FKR} 
  F.~Farakos, A.~Kehagias and A.~Riotto,
  {\it { {\small   On the Starobinsky Model of Inflation from Supergravity,}}}
  arXiv:1307.1137 [hep-th].
  %%CITATION = ARXIV:1307.1137;%%
  %12 citations counted in INSPIRE as of 26 Sep 2013
  
  \bibitem{FKLP12} 
  S.~Ferrara, R.~Kallosh, A.~Linde and M.~Porrati,
  {\it { {\small   Minimal Supergravity Models of Inflation,}}}
  arXiv:1307.7696 [hep-th].~ 
  %%CITATION = ARXIV:1307.7696;%%
  %8 citations counted in INSPIRE as of 26 Sep 2013 
  S.~Ferrara, R.~Kallosh, A.~Linde and M.~Porrati,
  {\it { {\small   Higher Order Corrections in Minimal Supergravity Models of Inflation,}}}
  arXiv:1309.1085 [hep-th].
  %%CITATION = ARXIV:1309.1085;%%
  %4 citations counted in INSPIRE as of 26 Sep 2013
  
  \bibitem{Fre} 
  P.~Fre and A.~S.~Sorin,
  {\it {{\small ``Inflation and Integrable one-field Cosmologies embedded in Rheonomic Supergravity,''}}}
  arXiv:1308.2332 [hep-th]. ~P. Fre, A. Sagnotti and A. S. Sorin, 
  {\it {{\small Integrable Scalar Cosmologies I. Foundations and links with
String Theory,}}} arXiv:1307.1910 [hep-th].~
E. Dudas, N. Kitazawa, S. P. Patil and A. Sagnotti, {\it {{\small CMB Imprints of a Pre-Inflationary Climbing
Phase,}}} JCAP 1205 (2012) 012, [arXiv:1202.6630 [hep-th]].~
A. Sagnotti,  {\it {{\small Brane SUSY Breaking and Inflation: Implications for Scalar Fields and CMB Distorsion,}}} arXiv:1303.6685 [hep-th].
  %%CITATION = ARXIV:1308.2332;%%
  %1 citations counted in INSPIRE as of 03 Oct 2013
  
  \bibitem{KS0} 
  S.~V.~Ketov and A.~A.~Starobinsky,
  {\it { {\small   Embedding $R+R^2$-Inflation into Supergravity,}}}
  Phys.\ Rev.\ D {\bf 83}, 063512 (2011)
  [arXiv:1011.0240 [hep-th]].
  %%CITATION = ARXIV:1011.0240;%%
  %23 citations counted in INSPIRE as of 26 Sep 2013
  
  \bibitem{KS1}
  S.~Ketov,
  {\it { {\small   F(R) supergravity,}}}
  AIP Conf.\ Proc.\  {\bf 1241} (2010) 613
  [arXiv:0910.1165 [hep-th]].~
   S.~V.~Ketov,
  {\it { {\small   Chaotic inflation in F(R) supergravity,}}}
  Phys.\ Lett.\ B {\bf 692}, 272 (2010)
  [arXiv:1005.3630 [hep-th]].~
   S.~V.~Ketov and S.~Tsujikawa,
  {\it { {\small   Consistency of inflation and preheating in F(R) supergravity,}}}
  Phys.\ Rev.\ D {\bf 86}, 023529 (2012)
  [arXiv:1205.2918 [hep-th]].~
  %%CITATION = ARXIV:1205.2918;%%
  %9 citations counted in INSPIRE as of 27 Sep 2013
  %%CITATION = ARXIV:1005.3630;%%
  %11 citations counted in INSPIRE as of 27 Sep 2013
  %%CITATION = ARXIV:0910.1165;%%
  %13 citations counted in INSPIRE as of 26 Sep 2013
  S.~V.~Ketov,
  {\it { {\small   On the supersymmetrization of f(R) gravity,}}}
  arXiv:1309.0293 [hep-th].
  %%CITATION = ARXIV:1309.0293;%%
  
  
\bibitem{FGN} 
  S.~Ferrara, M.~T.~Grisaru and P.~van Nieuwenhuizen,
  {\it { {\small   Poincare and Conformal Supergravity Models With Closed Algebras,}}}
  Nucl.\ Phys.\ B {\bf 138}, 430 (1978).
  %%CITATION = NUPHA,B138,430;%%
  %73 citations counted in INSPIRE as of 25 Sep 2013

  \bibitem{FKvP} 
  S.~Ferrara, R.~Kallosh and A.~Van Proeyen,
  {\it { {\small On the Supersymmetric Completion of $R+R^2$ Gravity and Cosmology,}}}
  arXiv:1309.4052 [hep-th].
  %%CITATION = ARXIV:1309.4052;%%
 
 \bibitem{ketov2}
  S.~V.~Ketov,
  {\it { {\small    On the supersymmetrization of f(R) gravity,}}}
  arXiv:1309.0293 [hep-th]. ~ S.~V.~Ketov and T.~Terada,
  {\it { {\small    Old-minimal supergravity models of inflation,}}}
  arXiv:1309.7494 [hep-th].
  %%CITATION = ARXIV:1309.0293;%%
 
 
 \bibitem{FvN} 
  S.~Ferrara and P.~van Nieuwenhuizen,
  {\it { {\small   The Auxiliary Fields of Supergravity,}}}
  Phys.\ Lett.\ B {\bf 74}, 333 (1978).
  %%CITATION = PHLTA,B74,333;%%
  %309 citations counted in INSPIRE as of 27 Sep 2013
  
 \bibitem{SW} 
  K.~S.~Stelle and P.~C.~West,
  {\it { {\small   Minimal Auxiliary Fields for Supergravity,}}}
  Phys.\ Lett.\ B {\bf 74}, 330 (1978).
  %%CITATION = PHLTA,B74,330;%%
  %301 citations counted in INSPIRE as of 27 Sep 2013
 
 \bibitem{FvP} 
  D.~Z.~Freedman and A.~Van Proeyen,
  {\it { {\small   Supergravity,}}}
  Cambridge, UK: Cambridge Univ. Pr. (2012) 607 p
 
\bibitem{ns1} 
  E.~Cremmer, S.~Ferrara, C.~Kounnas and D.~V.~Nanopoulos,
  {\it { {\small   Naturally Vanishing Cosmological Constant in N=1 Supergravity,}}}
  Phys.\ Lett.\ B {\bf 133}, 61 (1983).
  %%CITATION = PHLTA,B133,61;%%
  %566 citations counted in INSPIRE as of 26 Sep 2013
  J.~R.~Ellis, A.~B.~Lahanas, D.~V.~Nanopoulos and K.~Tamvakis,
  {\it { {\small   No-Scale Supersymmetric Standard Model,}}}
  Phys.\ Lett.\ B {\bf 134}, 429 (1984).
  %%CITATION = PHLTA,B134,429;%%
  %451 citations counted in INSPIRE as of 26 Sep 2013


  
  
%%%%%%%%%%%%%%%%%%%%%%%%%%%%%%%%%%%%%%%%%%%%%%%%%%%%%%%%%%%%%%%%%%%%%%
\end{thebibliography}
\end{document}